# Finite element simulation for validation of multi-dipole line cusp magnetic field configuration for MPD

A. D. Patel[1*], A. Amardas[1], N. Ramasubramanian[1,2]

[1]Institute for Plasma Research, Bhat, Gandhinagar, Gujarat-382428, India

[2]Homi Bhabha National Institute, Training School Complex, Anushaktinagar, Mumbai 400094, India

[*]E-mail: amitphy9898@gmail.com

**Abstract:**

A **M**ulti-dipole line cusp configured **P**lasma **D**evice (MPD) having six electromagnets with embedded Vacoflux-50 as a core material has been operated with a capability to experimentally control the field-free region, the radial profile of magnetic field, and pole magnetic field by changing magnet current. For the validation of multi-dipole line cusp magnetic field (MMF) configuration in a 3-D geometry, a finite element simulation has been performed using COMSOL software. This paper presents 3-D magnetic field simulation results of multi-pole line cusp magnetic field configuration performed over the 1.2m length and 40cm diameter chamber in the vacuum condition. The simulation results show good agreement with the experimentally measured magnetic field profile. The performed magnetic field simulation results clearly capture that this configuration has full control over a null region (nearly field-free region) as well is capable to change the magnetic field values and radial variation of the magnetic field. Moreover, the magnetic field profiles over the end cross section of the device have been discussed.

I. **Introduction:**

Plasma confined by multi-dipole cusp configurations has found fundamental applications in the evolution of negative/positive ion sources [1-4], material sciences, plasma-etching reactors [5], fundamental plasma physics, etc. The potentiality to confine a large volume of high-density uniform plasma is the main crucial feature of such configurations [6]. Although the studies of plasma confinement by multi-dipole cusp configuration emerged from nuclear fusion research, it is no more considered for fusion research since other substantial options are available. The negative/positive ion source with a multi-pole cusp configuration has demonstrated its capability for high current and stable ion beams which is essential for the new generation of accelerators [7]. Moreover, these types of multi-dipole cusp geometries are being applicable for basic plasma-physics studies (i.e. plasma waves and instability, plasma-particle interaction) in laboratory plasma devices due to their capability to confine large-volume uniform and quiescent plasmas [6]. Thus for the validation and future plasma modeling using COMSOL multi-physics, the 3-D finite element magnetic field simulation has been performed [8].

In this article, a simulation of multi-pole line cusp magnetic field configuration using COMSOL software tool for MPD having axial length l200 cm and 40 cm diameter [9, 10] has been discussed. The results obtained by performing simulation are compared with experimentally measured magnetic field values. The control over the null region, the control over radial variation of magnetic field values, and control over the pole magnetic field have been discussed. Moreover, in MPD authors have been trying to produce contact ionization Cesium plasma using coaxial tungsten hot plate-based cathode source [11]. This source has been placed at the end edge of the chamber. The tungsten filaments have been used to heat the tungsten plate. The ambient magnetic field cause J X B forces and scatter the electron-emitting tungsten filaments [12] and thus nearly field-free region of the multi-pole cusp magnetic field configuration plays a crucial role in placing the source. Thus the counter ($r$, $\theta$) plane of magnetic field configuration at the end edge of the device is crucial and has been discussed. This counter ($r$, $\theta$) plane of the multi-pole cusp magnetic field configuration also helps to choose the location of the hot tungsten plate-based Cesium plasma inside the null region cusp magnetic field configuration so that maximum axial plasma length can be achieved. The construction of the MPD device is described in section II, while section III contains the COMSOL simulation model and simulation results compared with the experimental values and the conclusion of the performed simulation results will be discussed in Section IV.

## II. Multi-pole cusp magnetic field device (MPD)

The schematic diagram of the complete setup of the MPD device is shown in figure 1 (a), figure 1 (b) represents the CATIA image, and figure (c) represents the real image. The device consists of a vacuum chamber with a pumping system, magnet system, a tungsten filament-based source for the plasma ignition, and basic plasma diagnostics. The main chamber of the MPD is made up of non-magnetic stainless steel-304. The diameter of this chamber is 40 cm and the axial length is 1500 cm. This chamber is attached with a Turbo-Molecular Pump (430 liters/sec) backed up by a rotary pump on the end side and is sufficient to produce a base pressure of $10^{-6}$ mBar. The six electromagnets are arranged in the multi-dipole cusp configuration along the circumference of a 40.5 cm diameter cylindrical chamber. Each electromagnet is wound with double pan-cake windings using hollow copper pipes, the cross-section of which is shown in figure 2 and figure 3 shows the real image of the electromagnet. More about electromagnets have been discussed in our earlier published article ref. 8 and 9. Rectangular Vacoflux-50 core material is embedded in each of the magnets as shown in figure 3. The Vacoflux-50 core is a ferromagnetic material made up of iron and cobalt in a ratio of 50:50. The Vacoflux-50 core material has relative permeability is > 4000, the saturation flux density is > 2 Tela, and the Curie temperature is 700 degree Celsius. These six electromagnets are powered by a set of power supplies connected in parallel.

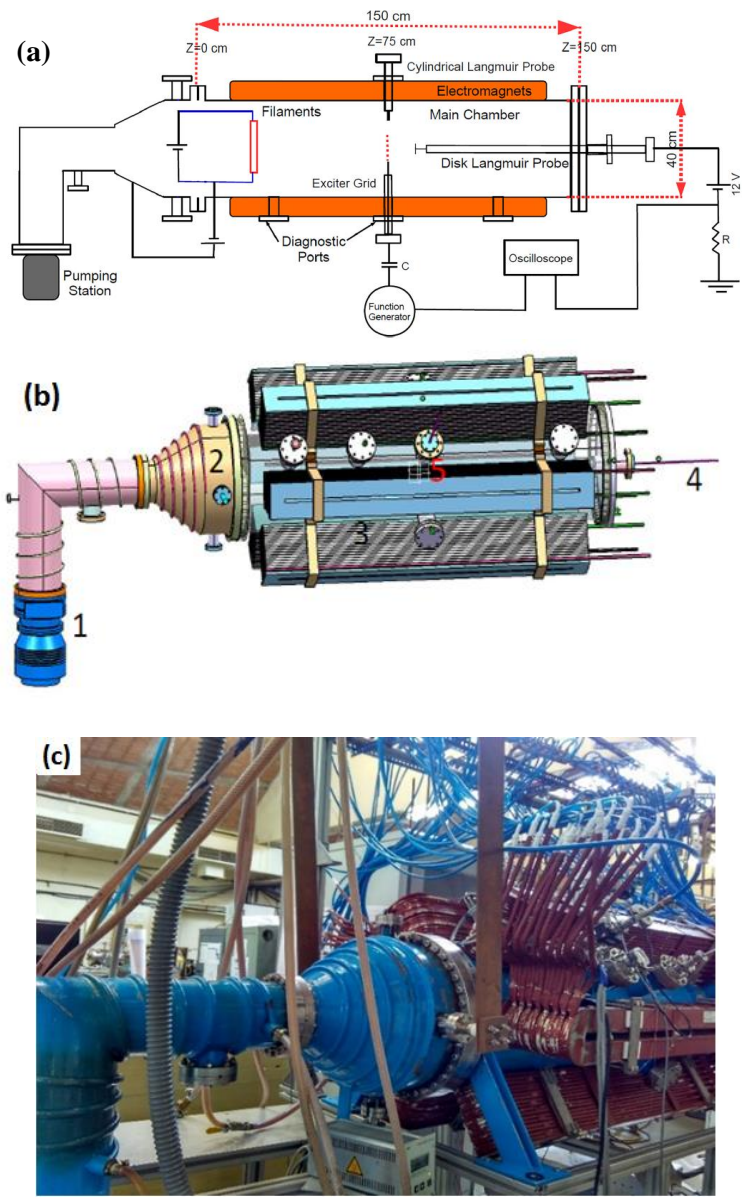

Fig. 1: (a) Schematic diagram of experimental setup of MPD (b) CATIA image of MPD (c) Image of MPD with versatile electro-magnets.

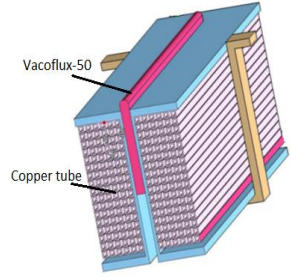

Fig. 2: Cross-sectional of the electromagnet showing copper pipes and core material vacoflux-50.

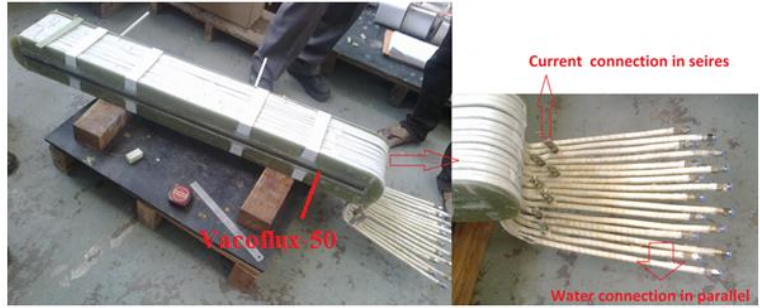

Fig. 3: Image of versatile electro magnet coil for multi-pole cusp magnetic field configuration.

### III. 3-D COMSOL magnetic field simulation and comparison with experimental results

A 3-D finite element magnetic field simulation for VMMF is performed using COMSOL software. Figure 4 shows the 3-D model created in COMSOL of the MPD device generated in COMSOL and figure 5 shows the Controlled fine mesh of the multi-pole cusp magnetic field configuration. Figure 6 represents the contour plot of the magnetic field lines at the mid ($r, \theta$) plane when 150A magnet current ($I_{mag}$) is pass through in magnets in vacuum condition. Similar simulations were also performed for 50A and 100A magnet current is passing through in magnets. The simulated magnetic field values along the cusp are compared with experimental results for these three currents. Figure 7 represents the radial variation of the magnetic field along the cusp region and is compared with experimental values at 50A, 100A, and 150A measured at the mid ($r, \theta$) plane of the device. The field inside the vacuum chamber is measured using a tri-axis gauss probe and a suitable gauss meter. The magnetic field measurements with changing magnet current were performed at several radial locations as well as on the different cross-sectional planes of the device. Figure 6 shows the 2-D ($r, \theta$) plane contour plot of the magnetic field simulated using COMSOL. The radial variation of the measured magnetic field ($B_M$) along the cusp region ($\theta = 0^o$) is shown in figure 4(a) for three different currents ($I_{mag}$= 50A and 100A, 150A) in the magnets. The figure also shows the simulated magnetic field ($B_S$) is in good agreement with measured values.

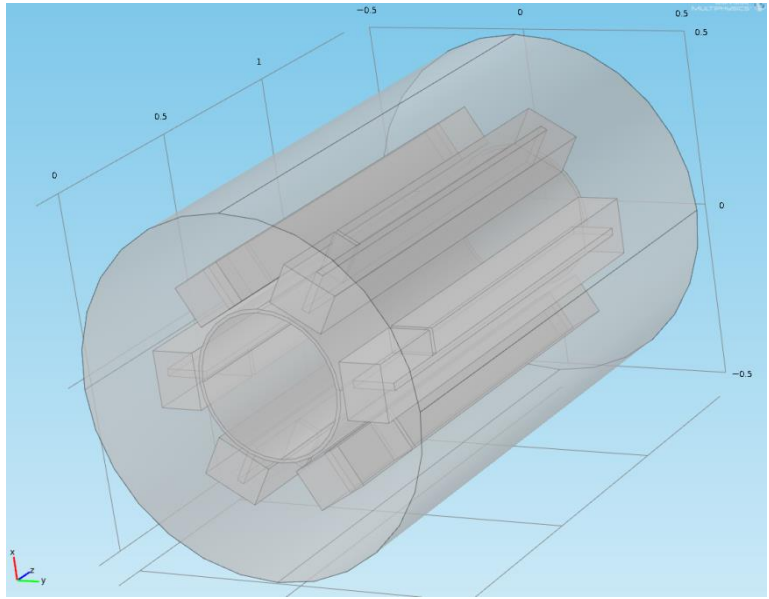

Fig. 4: A multi-dipole cusp magnetic field configuration model developed in COMSOL

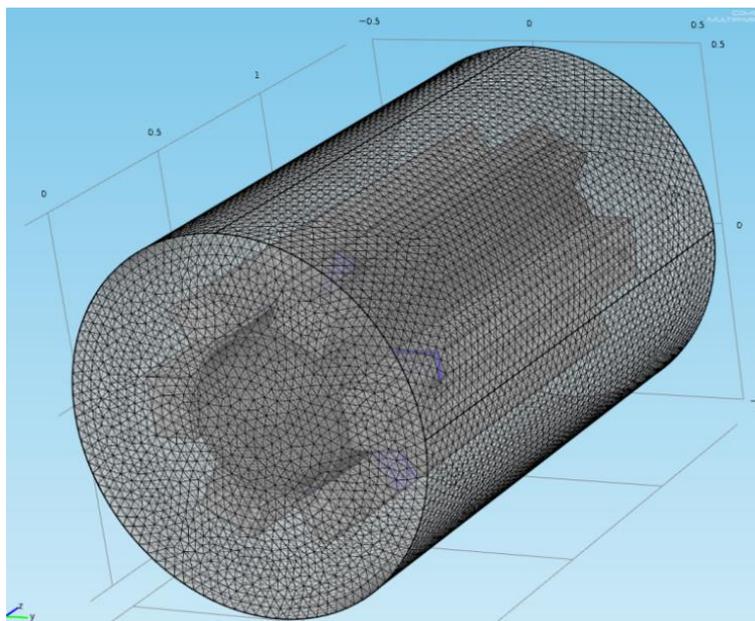

Fig. 5: Controlled fine mesh of the multi-pole cusp magnetic field configuration.

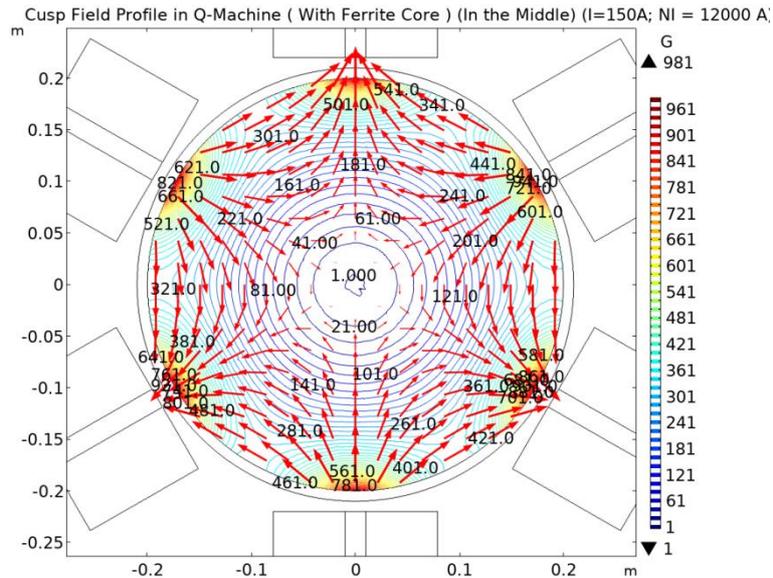
Fig. 6: Contour plot of the magnetic field lines in the mid (*r, θ*) plane, 150A magnet current, and inside the chamber of multi-pole cusp magnetic field configuration from COMSOL simulation

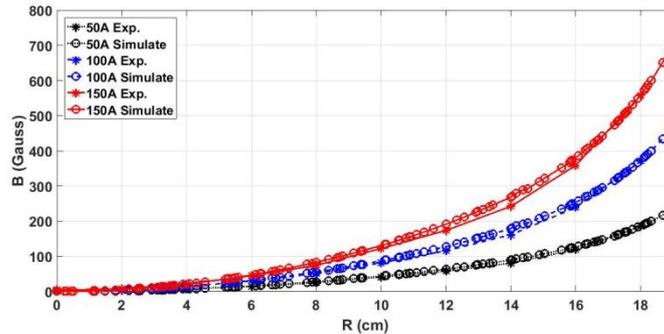
Fig. 7: Radial variation of magnetic field along cusp region and compared with experimental values at 50A, 100A, and 150A measured at mid (*r, θ*) plane of the device.

     A 3-D finite element magnetic field simulation using COMSOL software has been also performed for the end edge (*r, θ*) plane of the device. Figure 8 represents a contour plot of the vacuum field lines in the edge (*r, θ*) plane of the device at 150A magnet current of multi-pole cusp magnetic field configuration from COMSOL simulation. From the figure, it is clearly evident that the end edge (*r, θ*) plane follows the mid (*r, θ*) plane except for magnetic field values. Figure 9 represents the radial variation of the magnetic field along the cusp region at 50A, 100A, and 150A measured at the edge (*r, θ*) plane of the device in vacuum conditions. The variation of magnetic field values along the radial direction follows the trend of mid variation of magnetic field values along the radial direction of (*r, θ*) plane. The precise location of the end edges counter (*r, θ*) plane was obtained from the simulation has been used to choose the location of the hot tungsten plate-based cesium plasma source [11] inside the null region of cusp magnetic field configuration.

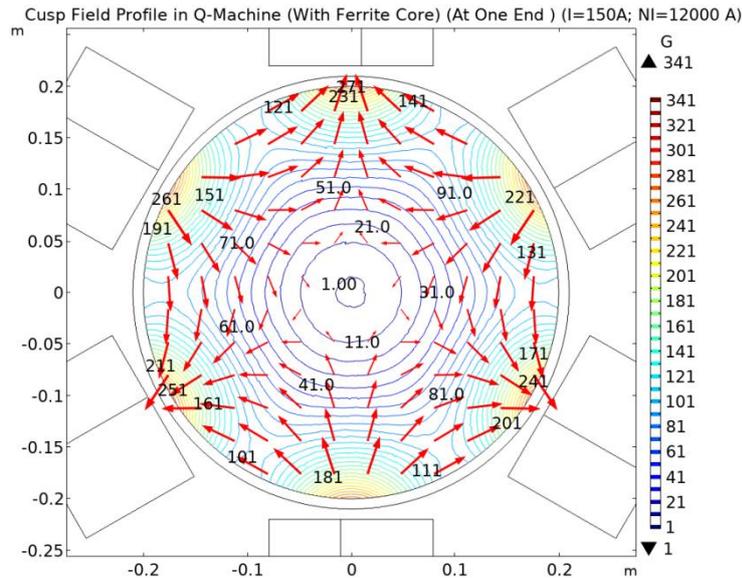

Fig. 8: Contour plot of the magnetic field lines in the edge (*r, θ*) plane, 150A magnet current, and inside the chamber of multi-pole cusp magnetic field configuration from COMSOL simulation

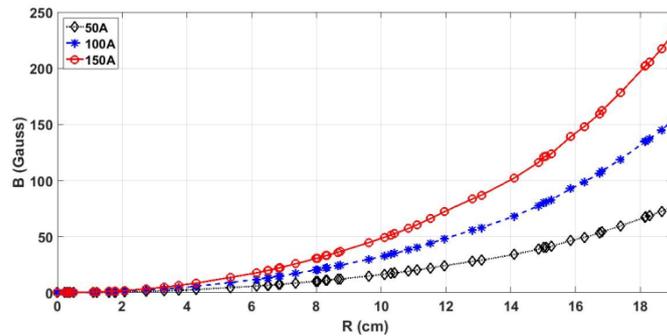

Fig. 9: Radial variation of magnetic field along cusp region at 50A, 100A, and 150A measured at edge (*r, θ*) plane of the device.

## IV. Conclusion

A Multi-dipole line cusp configured Plasma Device having six electromagnets with embedded Vacoflux-50 core material has been operated with a function to control the field free region, radial profile of magnetic field values, and pole magnetic field by changing magnet current experimentally. A simulation of a multi-dipole line cusp magnetic field (over a nearly 1.2 m axial length and 40 cm diameter of the chamber) in a vacuum is demonstrated using the COMSOL simulation tool. The simulation results show good agreement with experimentally measured magnetic field values. The performed magnetic field simulation results clearly capture that this configuration has full control over a null region (nearly field-free region) as well is capable to change the magnetic field values and radial variation of the magnetic field. Moreover, finite element magnetic field simulation using COMSOL software was also performed for observing the end edge (*r, θ*) plane of this configuration. The results show edge (*r, θ*) plane follows the mid (*r, θ*) plane trend except values and radial variation of magnetic field values also follows the trend except values.

**Acknowledgment:**